\documentclass[twocolumn,showpacs,amsmath,amssymb,superscriptaddress]{revtex4-1}
\usepackage{dcolumn,braket,color,amsmath,footnote,hyperref}
\usepackage{graphicx,bm,url,longtable}
\usepackage{tabularx,multirow,braket}
\newcommand\+{\dagger}
\newcommand\ft{\log{ft}}
\newcommand\MF{\braket{M_\text{F}}^2}
\newcommand\MGT{\braket{M_\text{GT}}^2}
\newcommand\MFb{\braket{M_\text{F}}}
\newcommand\MGTb{\braket{M_\text{GT}}}

\usepackage{longtable}	
\begin{document}

\title{$\beta$ decay of even-A nuclei within the interacting boson model based on nuclear density functional theory}

\author{K.~Nomura}
\email{knomura@phy.hr}
\affiliation{Department of Physics, Faculty of Science, University of
Zagreb, HR-10000 Zagreb, Croatia}

\author{R.~Rodr\'iguez-Guzm\'an}
\affiliation{Physics Department, Kuwait University, 13060 Kuwait, Kuwait}

\author{L.~M.~Robledo}
\affiliation{Departamento de F\'\i sica Te\'orica, Universidad
Aut\'onoma de Madrid, E-28049 Madrid, Spain}

\affiliation{Center for Computational Simulation,
Universidad Polit\'nica de Madrid,
Campus de Montegancedo, Bohadilla del Monte, E-28660-Madrid, Spain
}

\date{\today}

\begin{abstract}
We compute the $\beta$-decay $ft$-values within the frameworks of the 
energy density functional (EDF) and the interacting boson model (IBM). 
Based on the constrained mean-field calculation with the Gogny-D1M EDF, 
the IBM Hamiltonian for an even-even nucleus and essential ingredients 
of the interacting boson-fermion-fermion model (IBFFM) for describing 
the neighboring odd-odd nucleus are determined in a microscopic way. 
Only the boson-fermion and residual neutron-proton interaction 
strengths are determined empirically. The Gamow-Teller (GT) and Fermi 
(F) transition rates needed to compute the $\beta$-decay $ft$-values 
are obtained without any additional parameter or quenching of the $g_A$ 
factor. The observed $\ft$ values for the $\beta^+$ decays of the 
even-even Ba into odd-odd Cs nuclei, and of the odd-odd Cs to the 
even-even Xe nuclei, with mass $A\approx 130$ are reasonably well 
described. The predicted GT and F transition rates represent a 
sensitive test of the quality of the IBM and IBFFM wave functions. 
\end{abstract}


\maketitle


\section{Introduction}


The $\beta$ decay of atomic nuclei is a consequence of electro-weak 
fundamental processes and its ability to convert protons in neutrons 
and vice-versa make it a very relevant reaction mechanism in many 
nuclear physics scenarios. For instance, it plays an important role in 
modeling the creation of elements in astrophysical nucleosynthesis 
scenarios. Precise measurement and the theoretical description of the 
(single) $\beta$ decay are also crucial to better estimating the matrix 
element of the $\beta\beta$ decay, especially the one that does not 
emit neutrinos (neutrino-less $\beta\beta$ decay), a rare event that 
would signal the existence of physics beyond the Standard Model of 
elementary particles \cite{faessler1998,engel2017}. 

The quantitative understanding of the $\beta$ decay process requires a 
good and consistent description of the low-lying spectrum of both 
parent and daughter nuclei. A variety of theoretical methods have been 
used for this purpose. Without trying to be exhaustive, we can mention 
the quasiparticle random phase approximation (QRPA) used at various 
levels of sophistication 
\cite{sarriguren2001,simkovic2013,pirinen2015,marketin2016,mustonen2016,nabi2016}, 
the beyond mean-field approaches within the nuclear energy density 
functional (EDF) framework \cite{trrodriguez2010,yao2015}, the 
description based on large-scale interacting shell model calculations 
\cite{langanke2003,caurier2005,honma2005,shimizu2018,syoshida2018}, or 
that based on the interacting boson model (IBM) 
\cite{DELLAGIACOMA1989,navratil1988,IBFM,yoshida2002,zuffi2003,brant2004,brant2006,barea2009,yoshida2013,mardones2016}. 
Each of the theoretical methods have their own advantages and drawbacks 
that defined their range of applicability in the nuclear chart.

In the present work, we employ the IBM framework with input from 
microscopic EDF calculations \cite{nomura2008}. Our principal aim is a 
consistent theoretical description of the low-lying states and $\beta$ 
decay of even-A nuclei, including even-even and odd-odd ones. Within 
this approach, the potential energy surface (PES) for a given even-even 
nucleus is computed microscopically by means of the constrained 
mean-field method based on a nuclear EDF. The mean-field PES is mapped 
onto the expectation value of the IBM Hamiltonian in the intrinsic 
state of the $s$ (with spin and parity $0^+$) and $d$ ($2^+$) boson 
system. This procedure completely determines the strength parameters of 
the IBM Hamiltonian, which provides excitation spectra and transition 
strengths in arbitrary nuclear systems. 
The method can be extended to odd-mass and odd-odd nuclear systems by 
using the particle-boson coupling scheme. In such an extension, an 
additional  EDF mean-field calculation is carried out to provide the 
required spherical single-particle energies and occupation numbers for 
unpaired nucleon(s) in the odd-A or odd-odd nucleus. Those mean-field 
quantities represent an essential input to build the Hamiltonian of the 
interacting boson-fermion-fermion model (IBFFM) \cite{brant1984,IBFM}. 
The strength parameters for the boson-fermion and residual 
neutron-proton coupling terms are determined so as to reproduce 
reasonably well the experimental low-energy spectra in the neighboring 
odd-A nucleus, and the odd-odd nucleus of interest. At the price of 
having to determine these few coupling constants empirically, the 
method allows for a systematic, detailed and simultaneous description 
of spectroscopy in even-even, odd-A, and odd-odd nuclei in a 
computationally feasible manner as also required in $\beta$ decay studies. 

In Ref.~\cite{nomura2020beta}, we implemented the EDF-based interacting 
boson-fermion model (IBFM) \cite{iachello1979,IBFM} approach in the 
study of the $\beta$ decay of odd-A nuclei. There we studied the 
allowed $\beta$ decays, where the spin of the parent nucleus changes 
according to the $\Delta I=0,\pm 1$ rule and parity is conserved. Both 
the Gamow-Teller (GT) and Fermi (F) transition strengths were 
considered in the evaluation of  the $ft$ values of the $\beta$ decay. 
One of the advantages of our approach is that the calculation for the 
GT and F transition rates does not involve any additional free 
parameter, and that can be considered a very stringent test for the 
IBFM wave functions for the parent and daughter nuclei.

In the present work, since we are focusing in $\beta$ decay between 
even-mass nuclei, the calculation of the $ft$-values requires the IBM 
and IBFFM wave functions for the parent and daughter (or vise versa) 
even-even and odd-odd nuclei, respectively. Calculation of the 
single-$\beta$ decay between such even-A nuclei is also required for 
computing the $\beta\beta$ decay nuclear matrix elements, which are 
suggested to occur between a number of even-even nuclei. As in 
Ref.~\cite{nomura2020beta}, we consider the $\beta$ decays of those 
nuclei in the $A\approx 130$ mass region. There are a number of 
(phenomenological) IBFM calculations for the low-lying states and 
$\beta$ decay of odd-A nuclei. However, application of the IBFFM 
framework to the spectroscopy in odd-odd nuclear systems has rarely 
been pursued, let alone their $\beta$ decays. To the best of our 
knowledge, the IBFFM has been employed to study the $\beta$ decays only 
in Refs.~\cite{brant2006} and \cite{yoshida2013}. The former study is a 
first attempt to implement IBFFM in the $\beta$ decay of even-even 
system, but only one nucleus $^{124}$Ba was considered there. In the 
latter reference, two-neutrino $\beta\beta$ decays from Te to Xe 
isotopes was explored. The results of both studies are encouraging, 
since they show that the IBFFM framework is capable of describing the 
$\beta$ decay of even-A nuclei, even though the IBFFM Hamiltonian was 
determined in a fully phenomenological way. 

We have used the parametrization D1M \cite{D1M} of the Gogny-EDF 
\cite{Gogny,Rob19} for the microscopic calculation of the PES. Previous 
studies, using the EDF-to-IBM mapping procedure, have shown that the 
Gogny-D1M EDF provides a reasonable description of the spectroscopic 
properties of odd-A and odd-odd nuclei 
\cite{nomura2017odd-2,nomura2017odd-3,nomura2019dodd,nomura2020cs} in 
variety of mass regions including medium mass and heavy nuclei.

This paper is outlined as follows. In Sec.~\ref{sec:model} we briefly 
describe the procedures to build the IBFFM Hamiltonians from the 
constrained Gogny-EDF calculations, and introduce the $\beta$ decay 
operators. The results of our calculations for the low-lying energy 
levels of the even-even Xe and Ba, and odd-odd Cs nuclei are briefly 
reviewed in Sec.~\ref{sec:oo}. The $\ft$ values obtained for the 
$\beta$ decays of the studied even-A nuclei are discussed in 
Sec.~\ref{sec:beta}. Finally, Sec.~\ref{sec:summary} is devoted to the 
summary and the concluding remarks.


\section{Theoretical framework\label{sec:model}}


\subsection{Hamiltonian}

Let us first introduce the IBFFM Hamiltonian $\hat H$ for odd-odd systems. 
Note that we use the version of the IBFFM (called IBFFM-2)  
that distinguishes between neutron and proton 
degrees of freedom. 
The IBFFM-2 Hamiltonian consists of the IBM (called IBM-2) Hamiltonian $\hat H_\text{B}$ 
for an even-even nucleus, the Hamiltonians $\hat H_\text{F}^\rho$ for the odd neutron 
($\rho=\nu$) and proton ($\rho=\pi$), 
the Hamiltonians $\hat H^\rho_\text{BF}$ that couple the odd
neutron and the odd proton to the IBM-2 core, 
and finally the residual neutron-proton interaction $\hat V_\text{res}$: 
\begin{align}
\label{eq:ham}
 \hat H_\text{} = \hat H_\text{B} + \hat H_\text{F}^\nu + \hat
 H_\text{F}^\pi + \hat
  H_\text{BF}^\nu + H_\text{BF}^\pi + \hat V_\text{res}. 
\end{align}
The IBM-2 Hamiltonian reads: 
\begin{align}
\label{eq:ibm2}
 \hat H_{\text{B}} = \epsilon(\hat n_{d_\nu} + \hat n_{d_\pi})+\kappa\hat
  Q_{\nu}\cdot\hat Q_{\pi}
\end{align}
where $\hat n_{d_\rho}=d^\dagger_\rho\cdot\tilde d_{\rho}$ is the
$d$-boson number operator, and $\hat Q_\rho=d_\rho^\dagger s_\rho +
s_\rho^\dagger\tilde d_\rho^\dagger + \chi_\rho(d^\dagger_\rho\times\tilde
d_\rho)^{(2)}$ is the quadrupole operator. 
The parameters of the Hamiltonian are denoted by 
$\epsilon$, $\kappa$, $\chi_\nu$, and $\chi_\pi$. 
The doubly-magic nucleus $^{132}$Sn is taken as the inert 
core for the boson space. The numbers of neutron $N_{\nu}$ 
and proton $N_{\pi}$ bosons are computed as the numbers of 
neutron-hole and proton-particle pairs, respectively \cite{OAI}. 
In the following, we will simplify the notation and we will refer to the IBM-2 and IBFFM-2 
Hamiltonians simply 
as IBM and IBFFM, respectively.

The single-nucleon Hamiltonian $\hat H_\text{F}^\rho$ is given as 
\begin{align}
\label{eq:ham-f}
 \hat H_\text{F}^\rho = -\sum_{j_\rho}\epsilon_{j_\rho}\sqrt{2j_\rho+1}
  (a_{j_\rho}^\dagger\times\tilde a_{j_\rho})^{(0)}
\end{align}
with $\epsilon_{j_\rho}$ being the single-particle energy of the odd
nucleon. Here, $j_\nu$ ($j_\pi$) stands for the angular momentum of the single 
neutron (proton). The fermion creation and annihilation 
operators are denoted by $a_{j_\rho}^{(\dagger)}$ 
and $\tilde a_{j_\rho}$, with $\tilde a_{jm}=(-1)^{j-m}a_{j-m}$.
For the fermion valence space, we consider the full neutron and proton major
shell $N,Z=50-82$, i.e., the $3s_{1/2}$, $2d_{3/2}$, $2d_{5/2}$, $1g_{7/2}$, and $1h_{11/2}$ orbitals.

The boson-fermion coupling Hamiltonian $\hat H_\text{BF}^\rho$ 
takes the form: 
\begin{align}
\label{eq:ham-bf}
 \hat H_\text{BF}^\rho = \Gamma_\rho\hat Q_{\rho'}\cdot\hat q_{\rho} 
+
  \Lambda_\rho\hat V_{\rho'\rho} + A_\rho\hat n_{d_{\rho}}\hat n_{\rho}
\end{align}
where $\rho'\neq\rho$. 
The first, second, and third terms in the expression above are the
quadrupole dynamical, exchange, and monopole terms, respectively. 
The strength parameters are denoted by  
$\Gamma_\rho$, $\Lambda_\rho$, and $A_{\rho}$. 
As in previous studies 
\cite{scholten1985,arias1986} we assume that both the quadrupole dynamical and exchange terms 
are dominated by the interaction between unlike particles (i.e., between
the odd neutron and proton bosons and between the odd proton and neutron
bosons). On the other hand,  for the monopole term we only consider the interaction between
like-particles (i.e., between the odd neutron and neutron bosons and between
the odd proton and proton bosons). 
The bosonic quadrupole operator $\hat Q_\rho$ in Eq.~(\ref{eq:ham-bf}) 
has been defined in Eq.~(\ref{eq:ibm2}). 
The fermionic quadrupole operator $\hat q_\rho$ reads: 
\begin{align}
\hat q_\rho=\sum_{j_\rho j'_\rho}\gamma_{j_\rho j'_\rho}(a^\+_{j_\rho}\times\tilde
a_{j'_\rho})^{(2)},
\end{align} 
where $\gamma_{j_\rho
j'_\rho}=(u_{j_\rho}u_{j'_\rho}-v_{j_\rho}v_{j'_\rho})Q_{j_\rho
j'_\rho}$ and  $Q_{j_\rho j'_\rho}=\langle
l\frac{1}{2}j_{\rho}||Y^{(2)}||l'\frac{1}{2}j'_{\rho}\rangle$ represents
the matrix element of the fermionic 
quadrupole operator in the considered single-particle basis.
The exchange term $\hat V_{\rho'\rho}$ in Eq.~(\ref{eq:ham-bf}) reads: 
\begin{align}
\label{eq:Rayner-new-label}
 \hat V_{\rho'\rho} =& -(s_{\rho'}^\+\tilde d_{\rho'})^{(2)}
\cdot
\Bigg\{
\sum_{j_{\rho}j'_{\rho}j''_{\rho}}
\sqrt{\frac{10}{N_\rho(2j_{\rho}+1)}}\beta_{j_{\rho}j'_{\rho}}\beta_{j''_{\rho}j_{\rho}} \nonumber \\
&:((d_{\rho}^\+\times\tilde a_{j''_\rho})^{(j_\rho)}\times
(a_{j'_\rho}^\+\times\tilde s_\rho)^{(j'_\rho)})^{(2)}:
\Bigg\} + (H.c.), \nonumber \\
\end{align}
with $\beta_{j_{\rho}j'_{\rho}}=(u_{j_{\rho}}v_{j'_{\rho}}+v_{j_{\rho}}u_{j'_{\rho}})Q_{j_{\rho}j'_{\rho}}$.
In the second line of the expression above the notation $:(\cdots):$ indicates normal ordering. 
The definition of the number operator for the odd fermion 
in the monopole interaction has already appeared in Eq.~(\ref{eq:ham-f}).

For the residual neutron-proton interaction $\hat V_{\text{res}}$,  
we adopted the following form
\begin{align}
\label{eq:res}
 \hat V_{\text{res}}=4\pi u_{\rm D}\delta({\bf r})
+u_{\rm T}\Bigg\{\frac{3(\sigma_{\nu}\cdot{\bf r})
(\sigma_{\pi}\cdot{\bf r})}{r^2}-\sigma_{\nu}\cdot\sigma_{\pi}\Bigg\}, 
\end{align}
where the first and second terms denote the delta and tensor 
interactions, respectively. We have found that these two terms are
enough to provide a reasonable description 
of the low-lying states in the considered  odd-odd nuclei. 
Note that by definition ${\bf r}={\bf r_{\nu}}-{\bf r_{\pi}}$ and that 
$u_{\rm D}$ and $u_{\rm T}$ are the parameters of this term. 
Furthermore, the matrix element $V_\text{res}'$ of the residual interaction 
$\hat V_\text{res}$ can be expressed as
\cite{yoshida2013}: 
\begin{align}
\label{eq:vres}
V_\text{res}'
&= (u_{j_\nu'} u_{j_\pi'} u_{j_\nu} u_{j_\nu} + v_{j_\nu'} v_{j_\pi'} v_{j_\nu} v_{j_\nu})
V^{J}_{j_\nu' j_\pi' j_\nu j_\pi}
\nonumber \\
& {} - (u_{j_\nu'}v_{j_\pi'}u_{j_\nu}v_{j_\pi} +
 v_{j_\nu'}u_{j_\pi'}v_{j_\nu}u_{j_\pi}) \nonumber \\
&\times \sum_{J'} (2J'+1)
\left\{ \begin{array}{ccc} {j_\nu'} & {j_\pi} & J' \\ {j_\nu} & {j_\pi'} & J
\end{array} \right\} 
V^{J'}_{j_\nu'j_\pi j_\nu j_\pi'}, 
\end{align}
where
\begin{align}
V^{J}_{j_\nu'j_\pi'j_\nu j_\pi} = \langle j_\nu'j_\pi';J|\hat
 V_\text{res}|j_\nu j_\pi;J\rangle
\end{align}
represents the matrix element between the neutron-proton pair with 
angular momentum $J$. The bracket in Eq.~(\ref{eq:vres}) represents the 
corresponding Racah coefficient. As was done in 
Ref.~\cite{morrison1981}, the terms resulting from contractions are 
neglected in Eq.~(\ref{eq:vres}).

The matrix form of the IBFFM Hamiltonian of Eq.~(\ref{eq:ham}) is 
obtained in the basis $\ket{[L_{\nu}\otimes L_{\pi}]^{(L)}\otimes 
[j_{\nu}\otimes j_\pi]^{(J)}]^{(I)}} $. Here, $L_\rho$ is the angular 
momentum of proton or neutron boson system, $L$ is the total angular 
momentum of the boson system, and $I$ is the total angular momentum of 
the coupled boson-fermion-fermion system.

\subsection{Procedure to build the IBFFM Hamiltonian}

As the first step to build the IBFFM Hamiltonian, we have carried out 
(constrained) Hartree-Fock-Bogoliubov (HFB) calculations, based on the 
parametrization D1M of the Gogny-EDF. Those HFB calculations provide 
the potential energy surfaces (PESs), in terms of the quadrupole 
deformation parameters $\beta$ and $\gamma$, for the even-even core 
nuclei $^{124-132}$Xe and $^{124-132}$Ba. For a given nucleus, the 
Gogny-D1M PES is then mapped onto the expectation value of the IBM-2 
Hamiltonian in the boson coherent state \cite{ginocchio1980} (see, 
Refs.~\cite{nomura2008,nomura2010}, for details). This mapping 
procedure uniquely determines the parameters $\epsilon$, $\kappa$, 
$\chi_\nu$, and $\chi_\pi$ in the boson Hamiltonian. Their values have 
already been given in Ref.~\cite{nomura2020beta}.

Second, the single-particle energies $\epsilon_{j_\nu}$ 
($\epsilon_{j_\pi}$) and occupation probabilities $v^2_{j_\nu}$ 
($v^2_{j_\pi}$) of the unpaired neutron (proton) for the neighboring 
odd-N (odd-Z) nucleus are computed with the help of Gogny-D1M HFB 
calculations constrained to zero deformation \cite{nomura2017odd-2}. 
Those energies are used as input to the Hamiltonians $\hat 
H_\mathrm{F}^\nu$ ($\hat H_\mathrm{F}^\pi$) and $\hat 
H_\mathrm{BF}^\nu$ ($\hat H_\mathrm{BF}^\pi$), for the odd-$N$ Xe and 
odd-$Z$ Cs isotopes, respectively. The optimal values of the strength 
parameters for the boson-fermion Hamiltonian $\hat H_\mathrm{BF}^\nu$ 
($\hat H_\mathrm{BF}^\pi$), i.e., $\Gamma_\nu$, $\Lambda_\nu$, and 
$A_\nu$ ($\Gamma_\pi$, $\Lambda_\pi$, and $A_\pi$), are chosen 
separately for positive and negative parity, so as to reproduce the 
experimental low-energy spectrum for each of the considered odd-$N$ Xe 
(odd-$Z$ Cs) nuclei. 

Third, the use the previous strength parameters $\Gamma_\nu$, 
$\Lambda_\nu$, and $A_\nu$ ($\Gamma_\pi$, $\Lambda_\pi$, and $A_\pi$) 
for the IBFFM Hamiltonian for the odd-odd nuclei $^{124-132}$~Cs. In 
this case, the values of $\epsilon_{j_\rho}$ and $v^2_{j_\rho}$ are 
calculated again for each of these odd-odd nuclei. The employed 
boson-fermion interaction strengths, and $\epsilon_{j_\rho}$ and 
$v^2_{j_\rho}$ for the odd-odd Cs nuclei can be found in 
Ref.~\cite{nomura2020cs}. Finally, the strength parameters for the 
residual neutron-proton interaction, $u_\text{D}=0.7$ MeV and 
$u_\text{T}=0.02$ MeV, are taken also from Ref.~\cite{nomura2020cs}.

\subsection{Gamow-Teller and Fermi transition operators}

To obtain the $\beta$-decay $ft$-values, 
the Gamow-Teller (GT) and Fermi (F) 
matrix elements should be computed by using the IBM and IBFFM 
wave functions that correspond to the  
initial state (with spin $\ket{I_\mathrm{i}}$) for the parent nucleus 
and the final state (with spin $\ket{I_\mathrm{f}}$) for the daughter nucleus, 
or vice versa. 
The building blocks are the following one-fermion transfer operators 
\cite{DELLAGIACOMA1989}: 
\begin{align}
\label{eq:creation1}
A^{(j)\dagger}_{m} &= \zeta_{j} a_{jm}^{\dagger}
 + \sum_{j'} \zeta_{jj'} s^{\dagger}_\rho (\tilde{d}_\rho\times a_{j'}^{\dagger})^{(j)}_{m}\\
    \label{eq:creation2}
B^{(j)\dagger}_{m} &= \theta_{j} s^{\dagger}_\rho\tilde{a}_{jm}
 + \sum_{j'} \theta_{jj'} (d^{\dagger}_\rho\times\tilde{a}_{j'})^{(j)}_{m}
\end{align}
Both operators increase the number of valence 
neutrons (protons) $n_{j} + 2N_\rho$ by one. 
Note, that the index of $j_\rho$ is omitted 
for the sake of simplicity. 
The conjugate operators read:
\begin{align}
\label{eq:annihilation1}
\tilde{A}^{(j)}_{m}
= \zeta_{j}^{*} \tilde{a}_{jm}
+ \sum_{j'} \zeta_{jj'}^{*} s_\rho (d^{\dagger}_\rho\times\tilde{a}_{j'})^{(j)}_{m}\\
\label{eq:annihilation2}
\tilde{B}^{(j)}_{m}
= -\theta_{j}^{*} s_\rho a_{jm}^{\dagger}
- \sum_{j'} \theta_{jj'}^{*} (\tilde{d}_\rho\times a_{j'}^{\dagger})^{(j)}_{m}
\end{align}
These operators decrease the 
number of valence neutrons (protons) $n_{j} + 2N_\rho$ by one.

The coefficients $\zeta_{j}$, $\zeta_{jj'}$, $\theta_{j}$, and 
$\theta_{jj'}$ in Eqs.~(\ref{eq:creation1})-(\ref{eq:annihilation2}) 
are given \cite{IBFM} by \begin{align} \label{eq:zeta1} \zeta_{j} &= 
u_{j} \frac{1}{K_{j}'} , \\ \label{eq:zeta2} \zeta_{jj'} &= -v_{j} 
\beta_{j'j} \sqrt{\frac{10}{N_\rho(2j+1)}}\frac{1}{K K_{j}'} , \\ 
\label{eq:theta1} \theta_{j} &= \frac{v_{j}}{\sqrt{N_\rho}} 
\frac{1}{K_{j}''} , \\ \label{eq:theta2} \theta_{jj'} &= u_{j} 
\beta_{j'j} \sqrt{\frac{10}{2j+1}} \frac{1}{K K_{j}''} . \end{align} 
The parameters $K$, $K_j'$, and $K_j''$ entering the previous 
expressions read 
\cite{dellagiacoma1988phdthesis,DELLAGIACOMA1989,IBFM}: 
\begin{subequations} 
\begin{align} &K = \left( \sum_{jj'} 
\beta_{jj'}^{2} \right)^{1/2},\\ &K_{j}' = \left( 1 + 2 
\left(\frac{v_{j}}{u_{j}}\right)^{2} \frac{\braket{(\hat 
n_{s_\rho}+1)\hat n_{d_\rho}}_{0^+_1}} {N_\rho(2j+1)} \frac{\sum_{j'} 
\beta_{j'j}^{2}}{K^{2}} \right)^{1/2} ,\\ &K_{j}'' = \left( 
\frac{\braket{\hat n_{s_\rho}}_{0^+_1}}{N_\rho} 
+2\left(\frac{u_{j}}{v_{j}}\right)^{2} \frac{\braket{\hat 
n_{d_\rho}}_{0^+_1}}{2j+1} \frac{\sum_{j'} \beta_{j'j}^{2}}{K^{2}} 
\right)^{1/2} \end{align} 
\end{subequations} 
In these definition, $\hat n_{s_{\rho}}$ is the number operator for the 
$s_\rho$ boson and $\braket{\cdots}_{0^+_1}$ represents the expectation 
value of a given operator in the $0^+_1$ ground state of the considered 
even-even nucleus. For a more detailed account, the reader is referred
to Refs.~\cite{dellagiacoma1988phdthesis,DELLAGIACOMA1989,IBFM}.

The boson images of the Fermi ($t^\pm$) and Gamow-Teller 
($t^\pm\sigma$) transition operators, denoted by $\hat O^\text{F}$ and 
$\hat O^\text{GT}$, respectively, take the form
\begin{align}
&t^\pm\longmapsto\hat{O}^{\rm F} =-\sum_{j}\sqrt{2j+1} 
\left(P^{(j)}_{\nu}\times P^{(j)}_{\pi}\right)^{(0)}, \\
&t^\pm\sigma\longmapsto\hat{O}^{\rm GT} = \sum_{j'j}
 \eta_{j'j} \left(P^{(j')}_{\nu}\times P^{(j)}_{\pi}\right)^{(1)}
\end{align}
where the $\eta_{j'j}$ coefficients are proportional to the reduced matrix 
elements of the spin operator
\begin{align}
\label{eq:eta}
\eta_{j'j} 
&= - \frac{1}{\sqrt{3}} \langle\ell' \frac{1}{2} ; j' ||
 {\bf \sigma} || \ell \frac{1}{2} ; j\rangle
 \nonumber \\
&= - \delta_{\ell'\ell} \sqrt{2(2j'+1)(2j+1)}
W\left(\ell j' \frac{1}{2} 1 ; \frac{1}{2} j\right) ,
\end{align}
with $W$ being a Racah coefficient. In the case of  
$\beta^+$ decay $P_{\nu}^{(j')} = \tilde B_\nu^{(j')}$ 
and $P_{\pi}^{(j)} = \tilde A_\pi^{(j)}$ while 
for $\beta^-$ decay
$P_{\nu}^{(j')} = B_\nu^{(j')\dagger}$ 
and $P_{\pi}^{(j)} = {A_\pi^{(j)\dagger}}$. 
Then, the reduced Fermi $\MFb$ and Gamow-Teller 
$\MGTb$ matrix elements read: 
\begin{align}
&\MFb
=\frac{1}{\sqrt{2I_\mathrm{i}+1}} |\langle I_\mathrm{f} ||\hat{O}^\mathrm{F} || I_\mathrm{i} \rangle |\\
%
&\MGTb
=\frac{1}{\sqrt{2I_\mathrm{i}+1}} |\langle I_\mathrm{f} ||\hat{O}^\mathrm{GT} || I_\mathrm{i} \rangle |
\end{align}
The $ft$-value for the 
$\beta$ decay $I_\mathrm{i}\rightarrow  I_\mathrm{f}$, 
can be computed, in seconds, using the expression
\begin{align}
\label{eq:ft}
ft=\frac{6163}{\MF+g_A^2\MGT}.
\end{align}
The quantity $g_A$ is the ratio of the axial-vector to vector coupling constants, 
$g_A=G_{A}/G_{V}$. We have employed the free nucleon value  
$g_A=1.2701(25)$ \cite{beringer2012} for all the studied nuclei  
without quenching.


\section{Low-energy structure of the parent and daughter nuclei\label{sec:oo}}


\begin{figure*}[htb!]
\begin{center}
\includegraphics[width=0.7\linewidth]{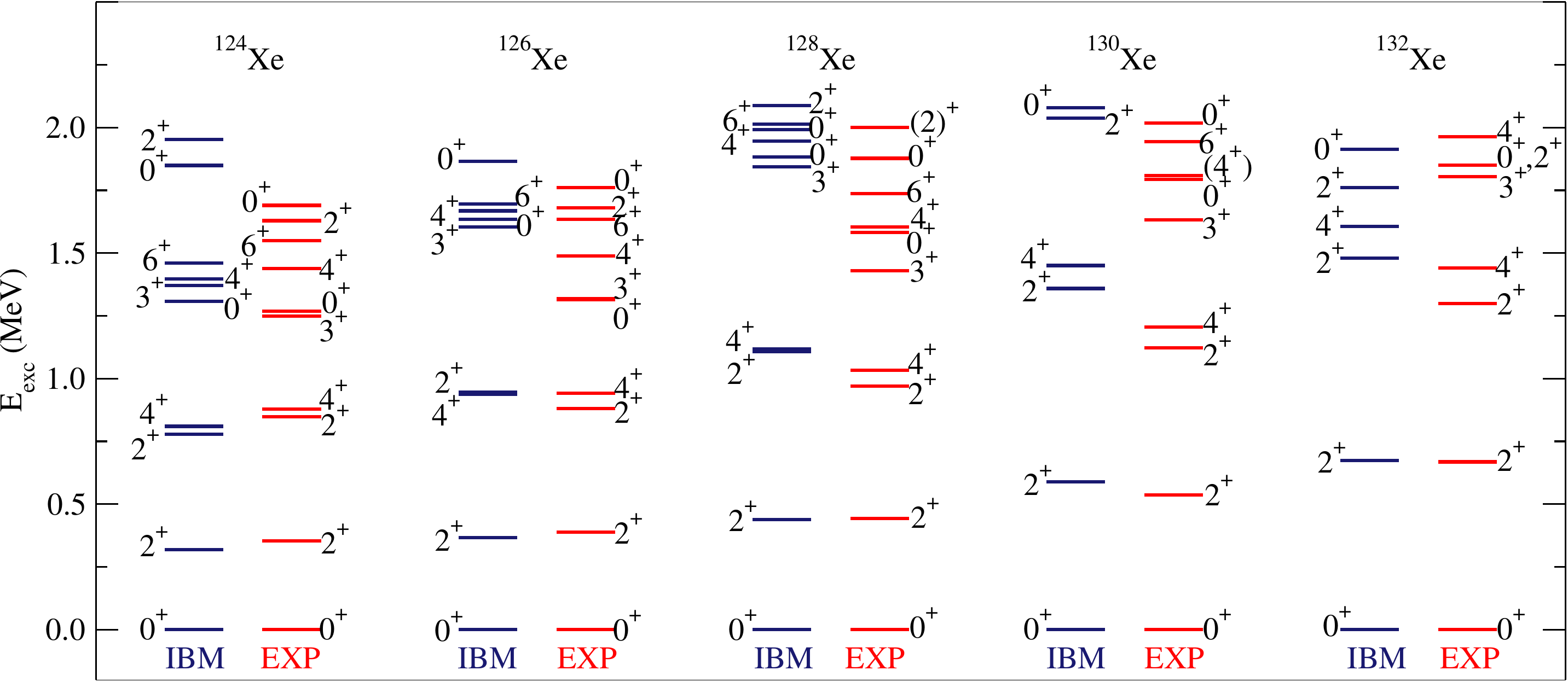}\\
\includegraphics[width=0.7\linewidth]{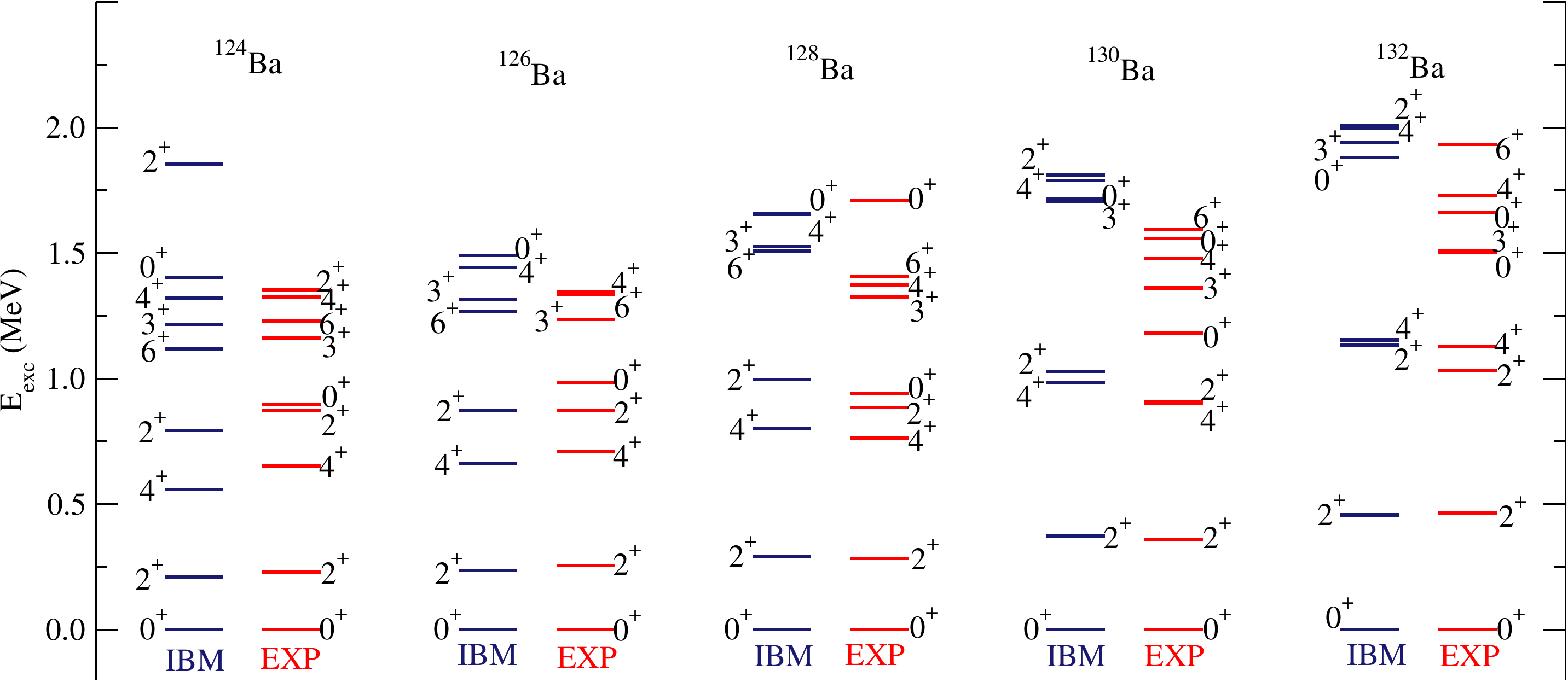}\\
\includegraphics[width=0.7\linewidth]{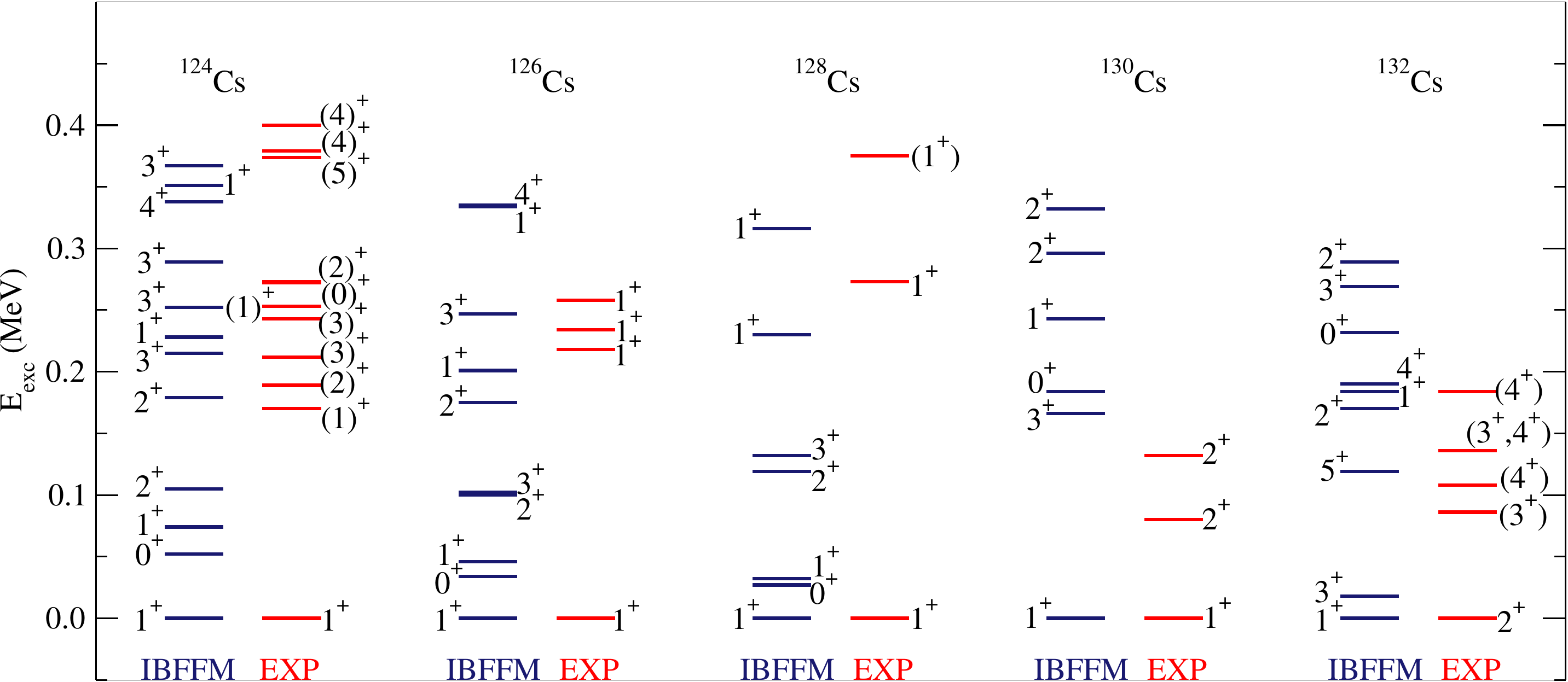}
\caption{(Color online) 
Low-energy and low-spin excitation spectra for the even-even $^{124-132}$Xe 
and $^{124-132}$Ba, and the odd-odd  $^{124-132}$Cs nuclei considered in this work. 
The experimental data are taken from the compilation of the ENSDF database \cite{data}.}
\label{fig:cs_level}
\end{center}
\end{figure*}

The low-lying level structure and electromagnetic properties 
of the even-even Ba and Xe, as well as odd-odd Cs, nuclei 
have already been discussed in detail in Ref.~\cite{nomura2020cs}. Also in this reference,  
the IBFM description of the neighboring odd-$N$ Ba and Xe 
as well as odd-$Z$ Cs isotopes has been amply discussed and 
shown to be also in reasonable agreement with experiment. 
The even-even $^{124-132}$Xe nuclei are taken as the 
cores for the odd-odd $^{124-132}$Cs nuclei. 
The Gogny-HFB PESs and mapped IBM description of these even-even Xe 
nuclei have shown that quite many of them are  $\gamma$-soft \cite{nomura2017odd-3}. 
Therefore, the low-energy spectra of the odd-odd Cs nuclei are described in terms of 
an unpaired neutron hole and a proton coupled to the 
$\gamma$-soft even-even core Xe nuclei. 
The low-energy positive-parity states of the odd-odd Cs nuclei, 
with excitation energies typically up to  $E_\text{exc}\approx 0.4$ MeV, 
have been shown \cite{nomura2020cs} to be built on the 
$(\nu sdg)^{-1}\otimes(\pi sdg)^1$ neutron-proton pair configuration. 

The low-energy spectra for the even-even $^{124-132}$Ba and 
$^{124-132}$Xe isotopes are shown in Fig.~\ref{fig:cs_level}. 
The IBM description of the low-lying excited states of the even-even systems 
looks very nice. An exception is perhaps the excitation energy 
of the $0^+_2$ level in many of the Ba nuclei, which is overestimated 
by the calculation. 
This is because the Gogny-HFB PESs have a minimum at a rather large  deformation 
leading to a rather pronounced rotational energy spectrum in the IBM calculation. 
Also in Fig.~\ref{fig:cs_level}, the calculated and experimental positive-parity 
excitation spectra for the odd-odd $^{124-132}$Cs nuclei 
are depicted up to 0.4 MeV excitation energy. 
Note that the energy spectra for the odd-odd Cs, shown in the figure, 
are taken from Ref.~\cite{nomura2020cs} 
without any modification. 
Except for the $^{132}$Cs nucleus, the IBFFM reproduces the correct 
ground-state spin. The calculation is not able to describe all the details 
of the lowest-lying level structures in each nucleus. 
But this is not surprising, considering the presence of both the unpaired 
neutron and proton degrees of freedom. 

There are also those higher-spin positive-parity states with an 
excitation energy of $E_\text{exc}\geqslant 0.4$ MeV and with a spin 
typically $I\geqslant 6^+$. These high-spin states are mostly accounted 
for by the $(\nu h_{11/2})^{-1}\otimes(\pi h_{11/2})^1$ configuration, 
and are  considered to be members of a chiral band. Note, however, that 
most of the $\beta$ decays considered in this work are relevant only 
for the low-spin states, i.e., $I\leqslant 4^+$, of the odd-odd Cs. 
Furthermore, the admixture of the $(\nu h_{11/2})^{-1}\otimes(\pi 
h_{11/2})^1$ pair configuration into the lowest-lying states is so 
small, that its effect on the $\beta$ decay rates from and to the 
ground states of the odd-odd Cs can be considered as negligible. 

Other spectroscopic properties of the low-lying states of the even-even 
Xe and Ba and odd-odd Cs isotopes, such as the electric quadrupole and 
magnetic dipole moments and transition strengths, were described 
reasonably well \cite{nomura2020cs}. 


\section{$\beta$ decay\label{sec:beta}}


\begin{figure}[htb!]
\begin{center}
\includegraphics[width=\linewidth]{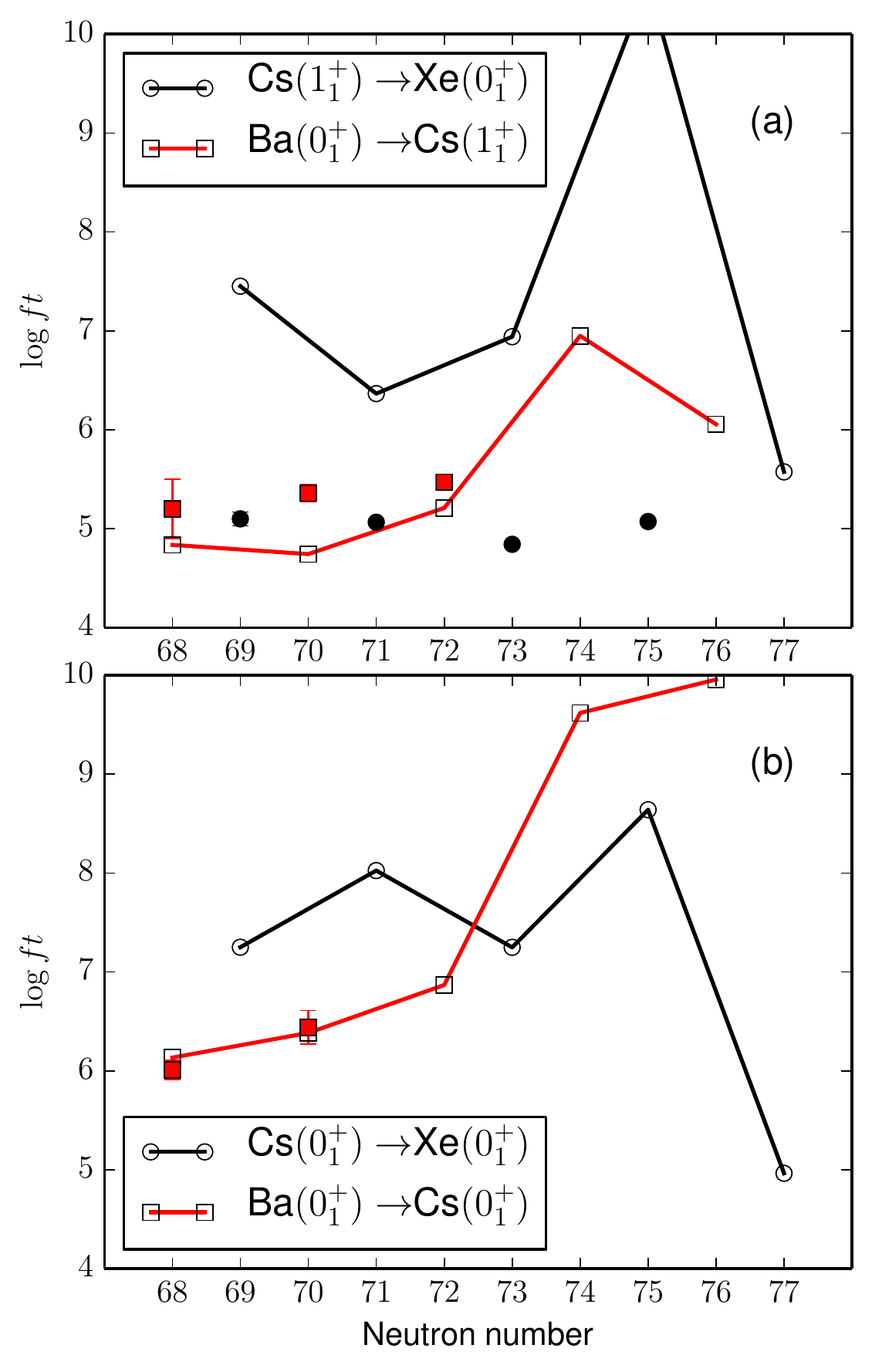}
\caption{(Color online) 
The predicted $\ft$ values (open symbols connected by solid lines) for 
the $\beta^+$ decays/EC (electron capture) between even-A nuclei, i.e., 
(a) Ba$(0^+_1)\rightarrow$Cs$(1^+_1)$, and 
Cs$(1^+_1)\rightarrow$Xe$(0^+_1)$, 
and 
(b) Ba$(0^+_1)\rightarrow$Cs$(0^+_1)$, and 
Cs$(0^+_1)\rightarrow$Xe$(0^+_1)$, 
plotted against the neutron number.  
The experimental $\ft$ values, taken from the ENSDF database \cite{data}, 
are represented by filled circles (Cs$\rightarrow$Xe decays) and squares 
(Ba$\rightarrow$Cs decays). 
}
\label{fig:logft}
\end{center}
\end{figure}

\subsection{$\beta$ decays of even-even Ba\label{sec:beta-bacs}}

\begin{table}[!htb]
\begin{center}
\caption{\label{tab:beta-bacs} 
The predicted and experimental $\log{ft}$ values in seconds 
for the $\beta^+$ decays/EC of the even-even 
$^{124,126,128}$Ba into odd-odd $^{124,126,128}$Cs nuclei.
The experimental data have been taken from Ref.~\cite{data}. 
}
\begin{ruledtabular}
 \begin{tabular}{cccc}
\multirow{2}{*}{Decay} &
\multirow{2}{*}{$I_\mathrm{i}\rightarrow  I_\mathrm{f}$} & 
\multicolumn{2}{c}{$\log{ft}$}\\
 \cline{3-4}
  & & Theory & Experiment \\
\hline
%
$^{124}$~Ba$\rightarrow ^{124}$~Cs
& $0^+_1\rightarrow 0^+_1$ & 6.134 & 6.01(10)\footnotemark[1] \\
& $0^+_1\rightarrow 1^+_1$ & 4.837 & 5.2(3) \\
& $0^+_1\rightarrow 1^+_2$ & 5.902 & 5.07(7) \\
& $0^+_1\rightarrow 1^+_3$ & 5.289 & 5.72(9)\\
& $0^+_1\rightarrow 1^+_4$ & 6.300 & 6.01(10)\footnotemark[1]\\
& $0^+_1\rightarrow 1^+_5$ & 4.600 & 6.04(9)\footnotemark[2]\\
& $0^+_1\rightarrow 1^+_6$ & 4.948 & 6.23(10)\footnotemark[3]\\
& $0^+_1\rightarrow 1^+_7$ & 5.176 & 6.83(16)\footnotemark[4]\\
& $0^+_1\rightarrow 1^+_8$ & 5.437 & 4.54(7)\footnotemark[5]\\
$^{126}$~Ba$\rightarrow ^{126}$~Cs
& $0^+_1\rightarrow 0^+_1$ & 6.348 & 6.44(17) \\
& $0^+_1\rightarrow 1^+_1$ & 4.744 & 5.36(9) \\
& $0^+_1\rightarrow 1^+_2$ & 4.823 & 6.36(12) \\
& $0^+_1\rightarrow 1^+_3$ & 5.392 & 5.49(5) \\
& $0^+_1\rightarrow 1^+_4$ & 5.624 & 6.44(17)\\
& $0^+_1\rightarrow 1^+_5$ & 4.538 & 5.18(6) \\
& $0^+_1\rightarrow 1^+_6$ & 6.028 & 5.12(7) \\
& $0^+_1\rightarrow 1^+_7$ & 8.944 & 5.08(8) \\
& $0^+_1\rightarrow 1^+_8$ & 10.810 & 4.54(7) \\
$^{128}$~Ba$\rightarrow ^{128}$~Cs
& $0^+_1\rightarrow 0^+_1$ & 6.868 &  \\
& $0^+_1\rightarrow 1^+_1$ & 5.201 & 5.471(9) \\
& $0^+_1\rightarrow 1^+_2$ & 4.858 & 8.26(11)\footnotemark[6] \\
& $0^+_1\rightarrow 1^+_3$ & 5.324 & 7.83(5)\footnotemark[7] \\
& $0^+_1\rightarrow 1^+_4$ & 7.823 & 5.57(4) \\
& $0^+_1\rightarrow 1^+_5$ & 4.776 & 7.83(9)\footnotemark[8] \\
& $0^+_1\rightarrow 1^+_6$ & 6.784 & 7.33(7)\footnotemark[9] \\
& $0^+_1\rightarrow 1^+_7$ & 8.595 & 6.68(6)
\footnotetext[1]{$I=(0,1)^+$ level at 272 keV in $^{124}$~Cs.}
\footnotetext[2]{$I=(1,2)^+$ level at 401 keV in $^{124}$~Cs.}
\footnotetext[3]{$I=(1^+,2^+)$ level at 404 keV in $^{124}$~Cs.}
\footnotetext[4]{$I=(1,2)^+$ level at 444 keV in $^{124}$~Cs.}
\footnotetext[5]{$I=(1,2)^+$ level at 557 keV in $^{124}$~Cs.}
\footnotetext[6]{$I=0^-,1$ level at 215 keV in $^{128}$~Cs.}
\footnotetext[7]{$I=0^-,1$ level at 230 keV in $^{128}$~Cs.}
\footnotetext[8]{$I=0^-,1$ level at 317 keV in $^{128}$~Cs.}
\footnotetext[9]{$I=0^-,1$ level at 359 keV in $^{128}$~Cs.}
 \end{tabular}
 \end{ruledtabular}
\end{center} 
\end{table}

We show in Fig.~\ref{fig:logft} the predicted $\ft$ values for the
$\beta^+$ decay (electron-capture (EC)) of the even-A nuclei, that is,  
$^{124-132}$~Ba$(0^+_1)\rightarrow^{124-132}$~Cs$(1^+_1)$, and 
$^{124-132}$~Cs$(1^+_1)\rightarrow^{124-132}$~Xe$(0^+_1)$ (panel (a)), 
and $^{124-132}$~Ba$(0^+_1)\rightarrow^{124-132}$~Cs$(0^+_1)$, and 
$^{124-130}$~Cs$(0^+_1)\rightarrow^{124-130}$~Xe$(0^+_1)$ (panel (b)). 
The experimental data are taken from the ENSDF database \cite{data} 
and are also depicted in the figure. 
The calculated $\ft$ values for the ground-state-to-ground-state decays 
$0^+_1$(Ba)$\rightarrow 1^+_1$(Cs) are in a very good agreement 
with the experiment. 
The $\ft$ values for the $1^+_1$(Cs)$\rightarrow 0^+_1$(Xe) decays 
are, however, considerably overestimated. 
The $0^+_1\leftrightarrows 1^+_1$ decays involve only the GT transitions, 
and the $\MGT$ values for these decays appear to be too small in our calculation. We will discuss below the reason for this small matrix elements. 
As for the $0^+_1\rightarrow 0^+_1$ decays (see, Fig.~\ref{fig:logft}(b)), 
where only the Fermi transition enters, the present calculation gives a 
very good description of the experimental $\ft$ values for the decays 
of $^{124,126}$Ba into $^{124,126}$Cs nuclei. 
A characteristic systematic trend of the predicted $\ft$ value is that 
for both the $\Delta I=\pm 1$ (Fig.~\ref{fig:logft}(a)) and $\Delta I=0$ (Fig.~\ref{fig:logft}(b))
decays it changes abruptly between $N=72$ and $N=76$ (for the Ba$\rightarrow$Cs decays) 
and between $N=73$ and $N=77$  (for the Cs$\rightarrow$Xe decays). 
Such an abrupt change in the $\ft$ values could reflect the evolution of 
nuclear structure between axially-symmetric and $\gamma$-soft deformed states. 
Accordingly, the structures of those wave functions for the boson-core Xe nuclei, 
as well as the neighboring odd-A Xe and Cs nuclei, are supposed to be 
very different between before and after the shape transition.

In Table~\ref{tab:beta-bacs}, we make a more detailed comparison 
between the predicted and experimental $\ft$ values for the $\beta^+$ 
decays of the  $0^+_1$ ground states of even-even $^{124,126,128}$Ba 
into those higher-lying states in odd-odd $^{124,126,128}$Cs nuclei. 
Experimentally, there are many degenerate levels in those nuclei, and 
their spin and/or parities are also not firmly established. This makes 
very difficult to establish a one-to-one correspondence between the 
calculated and experimental $\ft$ values. Those situations involving 
experimental degenerate levels are mentioned in the footnotes of 
Table~\ref{tab:beta-bacs}. The same rule applies to 
Tables~\ref{tab:beta-124cs}, \ref{tab:beta-126cs}, and 
\ref{tab:beta-128cs}, in the following Sec.~\ref{sec:beta-csxe}. Among 
other things, we emphasize that, as seen from 
Table~\ref{tab:beta-bacs}, a very good agreement between the predicted 
and experimental $\ft$ values is obtained for the $\beta^+$ decays to 
the $1^+_1$ ground states of the $^{124,126,128}$Cs nuclei. This 
situation is already shown in Fig.~\ref{fig:logft}(a). In addition, the 
observed $\ft$ values for the $0^+_1\rightarrow 0^+_1$ decays of 
$^{124,126}$~Ba, where the GT transition is forbidden and only the 
Fermi transition enters, are well reproduced by our calculation. As for 
the $\ft$ values for the decays into higher-energy, non-yrast $0^+$ and 
$1^+$ states, especially those higher than the 4-th lowest-energy ones 
for each spin, the discrepancy between the calculation and experiment 
increases. We can understand this trend by taking into account that the 
IBFFM is built in a reduced valence space, and the strength parameters 
are determined so as to fit the low-lying energy levels of odd-odd 
nuclei. Therefore, the wave functions for these higher non-yrast states 
in the present IBFFM calculation are not supposed to be as reliable as 
the ones of the lowest-energy states.

\subsection{$\beta$ decays of odd-odd Cs\label{sec:beta-csxe}}

\begin{table}[!htb]
\begin{center}
\caption{\label{tab:beta-124cs} 
Same as Table~\ref{tab:beta-bacs}, but for the $\beta^+$/EC decays 
of the odd-odd $^{124}$Cs to even-even $^{124}$Xe nuclei. 
}
\begin{ruledtabular}
 \begin{tabular}{cccc}
\multirow{2}{*}{Decay} &
\multirow{2}{*}{$I_\mathrm{i}\rightarrow  I_\mathrm{f}$} &  
\multicolumn{2}{c}{$\log{ft}$}\\
 \cline{3-4}
  & & Theory & Experiment \\
\hline
$^{124}$~Cs$\rightarrow ^{124}$~Xe
& $1^+_1\rightarrow 0^+_1$ & 7.452 & 5.10(7) \\
& $1^+_1\rightarrow 0^+_2$ & 6.659 & 5.54(7)\\
& $1^+_1\rightarrow 0^+_3$ & 6.621 & 6.21(7)\\
& $1^+_1\rightarrow 0^+_4$ & 7.560 & 5.72(7)\\
& $1^+_1\rightarrow 0^+_5$ & 7.496 & 5.69(7)\footnotemark[1]\\
& $1^+_1\rightarrow 0^+_6$ & 10.268 & 6.6(4)\footnotemark[2]\\
& $1^+_1\rightarrow 1^+_1$ & 5.810 & 6.6(4)\footnotemark[2]\\
& $1^+_1\rightarrow 1^+_2$ & 6.022 & 5.69(7)\footnotemark[1]\\
& $1^+_1\rightarrow 2^+_1$ & 6.061 & 5.10(7) \\
& $1^+_1\rightarrow 2^+_2$ & 8.949 & 5.89(7) \\
& $1^+_1\rightarrow 2^+_3$ & 8.628 & 5.73(7) \\
& $1^+_1\rightarrow 2^+_4$ & 7.017 & 6.15(7) \\
& $1^+_1\rightarrow 2^+_5$ & 7.245 & 6.01(7) \\
& $1^+_1\rightarrow 2^+_6$ & 6.636 & 6.43(7)\footnotemark[3]\\
& $1^+_1\rightarrow 2^+_7$ & 6.757 & 5.40(7)\\
& $1^+_1\rightarrow 2^+_8$ & 6.710 & 5.69(7)\footnotemark[1]\\
& $7^+_1\rightarrow 6^+_1$ & 6.965 & 7.4\\
& $7^+_1\rightarrow 6^+_2$ & 7.797 & 7.7\\
& $7^+_1\rightarrow 6^+_3$ & 10.596 & 7.6\\
& $7^+_1\rightarrow 6^+_4$ & 8.341 & 6.2\\
& $7^+_1\rightarrow 6^+_5$ & 8.862 & 7.3\footnotemark[4]\\
& $7^+_1\rightarrow 7^+_1$ & 7.761 & 7.6\\
& $7^+_1\rightarrow 7^+_2$ & 8.087 & 7.3\footnotemark[4]\\
& $7^+_1\rightarrow 7^+_3$ & 8.693 & 6.7\footnotemark[5]\\
& $7^+_1\rightarrow 7^+_4$ & 9.711 & 6.7\footnotemark[6]\\
& $7^+_1\rightarrow 8^+_1$ & 7.654 & 7.3\footnotemark[4]\\
& $7^+_1\rightarrow 8^+_2$ & 9.506 & 6.7\footnotemark[5]\\
& $7^+_1\rightarrow 8^+_3$ & 8.634 & 6.7\footnotemark[6]\\
\footnotetext[1]{$I=0^+,1^+,2^+$ at 2536 keV in $^{124}$~Xe.}
\footnotetext[2]{$I=(0^+,1,2)$ at 3897 keV in $^{124}$~Xe.}
\footnotetext[3]{$I=1^{(+)},2^{(+)}$ level at 2382 keV in $^{124}$~Xe}
\footnotetext[4]{$I=(6,7,8)^+$ level at 2979 keV in $^{124}$~Xe.}
\footnotetext[5]{$I=(6,7,8)^+$ level at 3739 keV in $^{124}$~Xe.}
\footnotetext[6]{$I=(6,7,8)^+$ level at 4093 keV in $^{124}$~Xe.}
 \end{tabular}
 \end{ruledtabular}
\end{center} 
\end{table}

\begin{table}[!htb]
\begin{center}
\caption{\label{tab:beta-126cs} 
Same as Table~\ref{tab:beta-bacs}, but for the $\beta^+$/EC decays 
of the odd-odd $^{126}$Cs to even-even $^{126}$Xe nuclei. 
}
\begin{ruledtabular}
 \begin{tabular}{cccc}
\multirow{2}{*}{Decay} &
\multirow{2}{*}{$I_\mathrm{i}\rightarrow  I_\mathrm{f}$} &  
\multicolumn{2}{c}{$\log{ft}$}\\
 \cline{3-4}
  & & Theory & Experiment \\
\hline
$^{126}$~Cs$\rightarrow ^{126}$~Xe
& $1^+_1\rightarrow 0^+_1$ & 6.366 & 5.066(19)\\ %
& $1^+_1\rightarrow 0^+_2$ & 6.832 & 5.39(3) \\ 
& $1^+_1\rightarrow 0^+_3$ & 7.345 & \\
& $1^+_1\rightarrow 0^+_4$ & 7.084 & 6.93\footnotemark[5]\\
& $1^+_1\rightarrow 0^+_5$ & 9.782 & 6.176(25)\footnotemark[6]\\
& $1^+_1\rightarrow 0^+_6$ & 7.907 & 5.941(25)\footnotemark[7]\\
& $1^+_1\rightarrow 0^+_7$ & 6.307 & 6.67(4)\footnotemark[8]\\
& $1^+_1\rightarrow 0^+_8$ & 6.465 & \\
& $1^+_1\rightarrow 0^+_9$ & 9.447 & 6.09(3)\footnotemark[9]\\
& $1^+_1\rightarrow 1^+_1$ & 6.161 & $>$7.1\footnotemark[4] \\
& $1^+_1\rightarrow 1^+_2$ & 7.639 & 6.93\footnotemark[5]\\
& $1^+_1\rightarrow 1^+_3$ & 8.279 & 6.176(25)\footnotemark[6]\\
& $1^+_1\rightarrow 2^+_1$ & 8.405 & 6.791(24) \\ 
& $1^+_1\rightarrow 2^+_2$ & 4.426 & 7.574(20) \\ 
& $1^+_1\rightarrow 2^+_3$ & 5.127 & 8.83(10) \\ 
& $1^+_1\rightarrow 2^+_4$ & 5.563 & 6.306(12) \\ 
& $1^+_1\rightarrow 2^+_5$ & 5.417 & 6.988(18) \\
& $1^+_1\rightarrow 2^+_6$ & 7.349 & $>$7.1\footnotemark[4] \\
\footnotetext[4]{$I=(1,2^+)$ level at 2215 keV in $^{126}$~Xe}
\footnotetext[5]{$I=0^+,1,2$ level at 2229 keV in $^{126}$~Xe}
\footnotetext[6]{$I=0^+,1,2$ level at 2347 keV in $^{126}$~Xe}
\footnotetext[7]{$I=0^+,1,2$ level at 2503 keV in $^{126}$~Xe}
\footnotetext[8]{$I=0^+,1,2$ level at 2521 keV in $^{126}$~Xe}
\footnotetext[9]{$I=0^+,1,2$ level at 2796 keV in $^{126}$~Xe}
 \end{tabular}
 \end{ruledtabular}
\end{center} 
\end{table}

\begin{table}[!htb]
\begin{center}
\caption{\label{tab:beta-128cs} 
Same as Table~\ref{tab:beta-bacs}, but for the $\beta^+$/EC decays 
of the odd-odd $^{128,130,132}$Cs to even-even $^{128,130,132}$Xe nuclei. 
}
\begin{ruledtabular}
 \begin{tabular}{cccc}
\multirow{2}{*}{Decay} &
\multirow{2}{*}{$I_\mathrm{i}\rightarrow  I_\mathrm{f}$} &  
\multicolumn{2}{c}{$\log{ft}$}\\
 \cline{3-4}
  & & Theory & Experiment \\
\hline
$^{128}$~Cs$\rightarrow ^{128}$~Xe
& $1^+_1\rightarrow 0^+_1$ & 6.941 & 4.843(10)\\ %
& $1^+_1\rightarrow 0^+_2$ & 7.236 & 5.579(24) \\ %
& $1^+_1\rightarrow 0^+_3$ & 8.783 & 7.48(4) \\ %
& $1^+_1\rightarrow 0^+_4$ & 6.744 & 5.70(3) \\ %
& $1^+_1\rightarrow 1^+_1$ & 5.853 & 6.37(3)\footnotemark[1]\\
& $1^+_1\rightarrow 1^+_2$ & 8.427 & 6.32(3)\footnotemark[2]\\
& $1^+_1\rightarrow 1^+_3$ & 7.577 & 6.49(3)\footnotemark[3]\\
& $1^+_1\rightarrow 2^+_1$ & 6.924 & 5.089(24)\\
& $1^+_1\rightarrow 2^+_2$ & 7.386 & 5.829(25)\\
& $1^+_1\rightarrow 2^+_3$ & 6.640 & 6.10(3)\\
& $1^+_1\rightarrow 2^+_4$ & 6.772 & 6.37(3)\footnotemark[1]\\
& $1^+_1\rightarrow 2^+_5$ & 7.130 & 6.24(3)\\
& $1^+_1\rightarrow 2^+_6$ & 6.867 & 6.32(3)\footnotemark[2]\\
& $1^+_1\rightarrow 2^+_7$ & 7.046 & 6.49(3)\footnotemark[3]\\
$^{130}$~Cs$\rightarrow ^{130}$~Xe
& $1^+_1\rightarrow 0^+_1$ & 10.523 & 5.073(6)\\ %
& $1^+_1\rightarrow 0^+_2$ & 6.334 & 7.0(1)\\ %
& $1^+_1\rightarrow 0^+_3$ & 6.711 & 6.2(1)\\ %
& $1^+_1\rightarrow 1^+_1$ & 6.800 & 6.9(2)\footnotemark[4]\\
& $1^+_1\rightarrow 1^+_2$ & 8.373 & \\
& $1^+_1\rightarrow 2^+_1$ & 7.309 & 6.3(1)\\
& $1^+_1\rightarrow 2^+_2$ & 5.233 & 7.5(4)\\
& $1^+_1\rightarrow 2^+_3$ & 4.826 & 6.2(1)\\
& $1^+_1\rightarrow 2^+_4$ & 7.564 & 6.9(2)\footnotemark[4]\\
$^{132}$~Cs$\rightarrow ^{132}$~Xe
& $1^+_1\rightarrow 0^+_1$ & 5.575 & \\ %
& $1^+_1\rightarrow 0^+_2$ & 6.991 & \\ %
& $1^+_1\rightarrow 2^+_1$ & 6.375 & \\ %
& $1^+_1\rightarrow 2^+_2$ & 4.388 & \\ %
& $2^+_1\rightarrow 2^+_1$ & 6.364 & 6.679\\
& $2^+_1\rightarrow 2^+_2$ & 4.459 & 8.7(1)\\
& $2^+_1\rightarrow 2^+_3$ & 4.685 & 6.61(2)\\
& $2^+_1\rightarrow 3^+_1$ & 6.367 & 7.17(2)\\
\footnotetext[1]{$I=1^+,2^+,3^+$ level at 2127 keV in $^{128}$~Xe.}
\footnotetext[2]{$I=(1,2^+)$ level at 2362 keV in $^{128}$~Xe.}
\footnotetext[3]{$I=(1,2^+)$ level at 2431 keV in $^{128}$~Xe.}
\footnotetext[4]{$I=1,2$     level at 2503 keV in $^{130}$~Xe.}
 \end{tabular}
 \end{ruledtabular}
\end{center} 
\end{table}

Experimental data is more abundant for the $\beta^+$ decay
of the odd-odd Cs isotopes. 
In Table~\ref{tab:beta-124cs}, the predicted $\ft$ values for the decays of the 
$1^+_1$ ground state of the $^{124}$Cs nucleus into $^{124}$Xe are compared 
with the experimental data.  
The $\ft$ value for the decay to the $I_f=0^+_1$ ground state 
of $^{124}$Xe is predicted to be unexpectedly larger than 
the experimental one. 
This means that the calculated GT transition rate is too small. 
The dominant pair configurations in the IBFFM wave function 
for the $1^+_1$ ground state are: 
$[(\nu s_{1/2})^{-1}\otimes (\pi s_{1/2})^1]^{(J=1^+)}$ (14.2 \%), 
$[(\nu s_{1/2})^{-1}\otimes (\pi d_{3/2})^1]^{(J=1^+)}$ (10.8 \%), 
$[(\nu s_{1/2})^{-1}\otimes (\pi d_{5/2})^1]^{(J=3^+)}$ (12.8 \%), and 
$[(\nu s_{1/2})^{-1}\otimes (\pi g_{7/2})^1]^{(J=3^+)}$ (11.2 \%). 
The rest of the wave function is made up of numerous other 
components that are so small in their magnitudes as to be neglected. 
The odd neutron occupying the 
$3s_{1/2}$ orbital makes dominant contributions to the above mentioned 
pair configurations. 
This seems to be consistent with the fact that 
the ${1/2}^+_1$ ground state of the neighboring odd-A nucleus 
$^{123}$Xe is mostly accounted for by the $3s_{1/2}$ neutron 
hole coupled to the even-even core $^{124}$Xe \cite{nomura2020beta}. 

Major contributions to the $\MGTb$ value for the 
$1^+_1\rightarrow 0^+_1$ decay of $^{124}$Cs turn out to be from 
those terms proportional to $[a^\+_{j_\nu}\times a^\+_{j_\pi}]^{(1)}$, 
$[\tilde d_\nu\times [a^\+_{j_\nu}\times a^\+_{j_\pi}]]^{(1)}$, 
and $[\tilde d_\nu\times\tilde d_\pi]\times[a^\+_{j_\nu}\times a^\+_{j_\pi}]]^{(1)}$. 
Their matrix elements are calculated as $0.02886$, $-0.02576$, and $-0.02152$, respectively. 
Two of these matrix elements have almost the equal magnitude but with opposite sign, 
and this leads to the too small $\MGTb$ value. 
A similar kind of cancellation, among different components 
in $\MGTb$ seems to take place in the GT transitions to the 
non-yrast $0^+$ states of the daughter nucleus. 
In the previous IBFFM calculation for the $\beta$ decay of $^{124}$Ba 
in Ref.~\cite{brant2006}, the same kind of discrepancy (too large 
$\ft$-values) was also found. 
The authors of Ref.~\cite{brant2006} also attributed it to 
the cancellations of small components in the GT matrix element. 
On the other hand, for the decays with $\Delta I=0$, i.e., 
$1^+_1\rightarrow 1^+_1$, the predicted $\ft$ values are 
generally smaller and reproduce the experimental data better 
than for the $\Delta I=\pm 1$ decays. 
This is mainly because $\MFb$ matrix elements also appear in 
the denominator in the expression for the $ft$-value 
for the $1^+_1\rightarrow 1^+_1$ decays (see, Eq.~(\ref{eq:ft}).

In the case of the $^{124}$Cs nucleus in particular, 
there are also the 
experimental data for the decays from the higher-spin state 
with $I_\text{i}=7^+$. 
In the present IBFFM calculation, those states with spin higher than 
$I\approx 7^+$ are formed mainly of the $(\nu h_{11/2})^{-1}\otimes (\pi h_{11/2})^1$ 
neutron-proton pair configuration. 
In contrast, the low-spin and low-energy states in the vicinity of the ground state and 
with $I\leqslant 4^+$ are mainly based on the neutron and 
proton positive-parity $sdg$ orbitals. 
The experimental $\ft$ values for the decays of $7^+$ state are 
generally $\ft\approx 7$, being larger than for the decays of the 
ground state $1^+_1$, which are typically $\ft\approx 5\sim 6$. 
The description of the $\ft$ values for this type of the $\beta$ decay 
in the present calculation appears to be good, at least for the 
decays to the few lowest levels in the daughter nucleus. 
We confirmed that, as expected, the configurations that involve 
the neutron and proton unique-parity $h_{11/2}$ single-particle 
orbitals make significant contributions to the $\MGTb$ matrix 
elements for the decays of the $I_\text{i}=7^+$ state.

The calculated $\ft$ for the rest of the odd-odd Cs nuclei are 
shown in Table~\ref{tab:beta-126cs} and Table~\ref{tab:beta-128cs} 
and compared to the experimental data. 
Similarly to the $^{124}$Cs $\rightarrow^{124}$Xe decay, the calculated 
$\ft$ values for the decays of $^{126-132}$~Cs, 
especially for the $\Delta I=\pm 1$ decays, are too large as compared with
the experimental values. This appears to happen mainly because 
the cancellation between components of the $\MGTb$ matrix elements 
occurs to the extent that too small GT rates are obtained. On the other hand, 
the agreement with experimental data is relatively good for the $\Delta I=0$ decays. 
Note, that a extremely large deviation is found in the decay 
$^{130}$~Cs$(1^+_1)\rightarrow^{130}$~Xe$(0^+_1)$ 
(see, Table~\ref{tab:beta-128cs} and also Fig.~\ref{fig:logft}). 
The predicted $ft$-value for this decay is larger than the experimental one 
implying a difference in the half-life of 5 orders of magnitude. 
Regarding the decay of the $^{132}$Cs nucleus, where experimentally 
the $2^+_1$ state is suggested to be the ground state, 
the present calculation reproduces 
very nicely the $\ft$ value for the  $2^+_1\rightarrow 2^+_1$ decay.


\section{Summary and concluding remarks\label{sec:summary}}

To summarize, the $\beta$ decays of even-$A$ nuclei are investigated 
within the EDF-based IBM approach. The even-even boson-core 
Hamiltonian, and essential building blocks of the particle-boson 
coupling Hamiltonians, i.e., single-particle energies and occupation 
probabilities of odd particles, are determined based on a 
fully-microscopic mean field calculation with the Gogny EDF. A few 
coupling constants for the boson-fermion Hamiltonians, and for the 
residual neutron-proton interaction, remain as the only free parameters 
of the model. They are determined as to reasonably reproduce the 
low-energy levels of each of the neighbouring odd-$A$ and odd-odd 
nuclei. The IBM and IBFFM wave functions obtained after diagonalization 
of the corresponding Hamiltonians for the parent and daughter nuclei 
are used to calculate Gamow-Teller and Fermi matrix elements, which are 
required to compute the $\beta$-decay $ft$-values. No additional 
parameter is introduced for the computations of the $ft$-values. The  
$g_A$ factor for the GT transition is not quenched in the present 
calculation. 

The present work, as well as the preceding ones of 
Refs.~\cite{nomura2020cs,nomura2020beta}, show that the employed 
theoretical method provides an excellent descriptions of the low-lying 
levels and electromagnetic transition rates in the relevant even-even 
Ba and Xe and odd-odd Cs, we as well as the neighboring odd-A Ba, Xe, 
and Cs nuclei. The observed $\beta^+$ decays/electron captures of the 
ground state $0^+_1$ of the even-even $^{124,126,128}$Ba into the 
$1^+_1$ ground states and $0^+_1$ states of $^{124,126,128}$Cs are 
described very nicely. On the other hand, the $\ft$ values for the 
$1^+\rightarrow 0^+$ decays of the odd-odd Cs into even-even Xe nuclei 
are, in general, predicted to be too large with respect to the 
experimental values. It is shown that the deviation occurs mainly 
because numerous small components in the GT matrix element $\MGTb$ 
cancel each other, leading to the too small GT transition rates. This 
is attributed to the combination of various factors in the adopted 
theoretical procedure, such as, the chosen boson-fermion coupling 
constants, residual neutron-proton interaction, and underlying 
microscopic inputs from the Gogny EDF.

The EDF-based IBM framework, employed here and in 
Ref.~\cite{nomura2020beta}, represents a computationally feasible 
theoretical method for studying simultaneously the low-energy 
excitations and fundamental processes in atomic nuclei. Especially, we 
consider the results for a single $\beta$ decay between even-mass 
nuclei quite encouraging, as they represent a crucial step toward the 
computation of the $\beta\beta$ decay nuclear matrix elements between 
even-even nuclei. Work along this direction is in progress and will be 
reported in near future.

\acknowledgments{}
The work of KN is financed within the Tenure Track Pilot Programme of 
the Croatian Science Foundation and the \'Ecole Polytechnique 
F\'ed\'erale de Lausanne, and the Project TTP-2018-07-3554 Exotic 
Nuclear Structure and Dynamics, with funds of the Croatian-Swiss 
Research Programme. The  work of LMR was supported by Spanish Ministry 
of Economy and Competitiveness (MINECO) Grants No. 
PGC2018-094583-B-I00.

\bibliographystyle{apsrev4-1}
\bibliography{refs}

\end{document}